\documentclass[12pt]{article}%
\usepackage{amssymb}
\usepackage{amsmath}
\usepackage{amsfonts}
\usepackage{graphicx}%
\providecommand{\U}[1]{\protect\rule{.1in}{.1in}}

\begin{document}
\bigskip%
\begin{titlepage}
\vspace{.3cm} \vspace{1cm}
\begin{center}
\baselineskip=16pt \centerline{\bf{\Large{ Unification of Gauge and Gravity}}}%
\baselineskip=16pt \centerline{\bf{\Large{Chern-Simons Theories in 3-D Space time }}}
\vspace{2truecm}
\baselineskip=16pt  \centerline{\large\bf Chireen A. Saghir, \ Laurence W. Shamseddine } \vspace{.5truecm}
\emph{\centerline{Physics Department, American University of Beirut, Lebanon}}
\end{center}
\vspace{2cm}
\begin{center}
{\bf Abstract}
\end{center}
Chamseddine and Mukhanov showed that gravity and gauge theories could be unified in one geometric construction provided that a metricity condition is imposed on the vielbein. In this paper we are going to show that by enlarging the gauge group we are able to unify Chern-Simons gauge theory and Chern-Simons gravity in 3-D space-time. Such unification leads to the quantization of the coefficients  for both Chern-Simons terms for compact groups but not for non-compact groups. Moreover, it leads to a topological invariant quantity of the 3 dimensional space-time manifold on which they are defined.
\end{titlepage}%

\section{Introduction}
Topological Field Theories (TFTs) are metric independent theories that define topological invariants of the manifold M. In general, we can distinguish between a topological field theory of Schwarz type and of Witten or cohomological type. In Schwarz type theories the action is explicitly independent on the metric\\
\begin{equation}
\frac{\delta S}{\delta g_{\mu \nu}} = 0  \label{1}%
\end{equation} 
Examples of such theories are Chern-Simons theories and BF theories. The other type is Witten type topological field theories like the Donaldson-Witten theory which was formulated by Witten in 1988 $\cite{Donaldson}$ and is used to compute Donaldson invariant. 

In a ground breaking paper $\cite{Jones}$, Witten (1989) showed that Chern-Simon gauge theories provide a physical description of a wide class of invariants of 3-manifolds and of knots and links in this manifold. The Chern-Simon action with a compact and simple gauge group G = SU(N)  on a generic 3-manifold is defined by%
\begin{equation}
S= \frac{k}{4 \pi} \int Tr ( A \wedge dA + \frac{2}{3} A \wedge A \wedge A), \label{1}%
\end{equation} 
Here k is the coupling constant and A is a G-gauge connection on the trivial bundle on M.\\
The partition function%
\begin{equation}
Z(M)=  \int DA e^{iS}, \label{2}%
\end{equation} 
is really a topological invariant of the manifold M as was proved by Witten $\cite{Jones}$. Moreover, if L is a collection of oriented and non-intersecting knots $C_i$, we have the partition function%
\begin{equation}
Z(M,L)=  \int DA  exp(iS) \prod_{i=1}^{r} W_{R_i}(C_i) \label{3}%
\end{equation}
Where\\
\begin{equation}
 W_R(C) = Tr_R P exp \int_{C} A \label{4}%
\end{equation}
represent a certain type of Knot invariant called Jones Polynomial.\\
\\
 In general, Chern Simons theories in 3-D exist for both gauge theory and gravity. Several work have been published to investigate the importance of Chern-Simons theories. In $\cite{deser}$, Deser, Jackiw and Templeton had shown that adding a topological Chern-Simons term to both three dimensional Yang-Mills and Gravity theories generate masses of the correspondind gauge particles and alters its content. Moreover, in $\cite{CHFro}$ Chamseddine and Frohlich showed that the lorentz and mixed lorentz-Weyl anomaly, but not the pure Weyl anomaly, of the two dimensional chiral bosons and fermions can be cancelled by the anomalies of the three dimensional gravitational Chern-Simons action. Quantum Gravity in 2+1 dimensions has been also studied $\cite{carlip}$. In a recent work $\cite{ChMu}$, Chamseddine and Mukhanov had shown that by enlarging the tangent group to SO(1,13) we are able to unify gravity and gauge interactions using a procedure that is applicable for an arbitrary dimension of the tangent space.\\
In this letter we show how to unify gauge and gravity Chern-Simons theories in 3-D space time. Such unification has several interesting results. First, we will be able to investigate the quantization of the coupling constants for both theories (gauge and gravity) with Lorentzian and Euclidean Signature. Second, we will obtain the resulting topological invariant quantity of the 3-D space-time manifold. \\

Before presenting our work, we would like to review the work of Chamseddine and Mukhanov $\cite{ChMu}$. In this work, it has been shown that it is possible to unify Yang-Mills gauge theories with gravity in terms of higher dimensional gauged Lorentz groups of the tangent space. The 4 dimensional manifold is spanned by the coordinate basis $e_{\alpha}$ where $\alpha$ = 1,…..4, and the $N \geqslant 4$- dimensional tangent space is spanned by $v_A$, where A = 1, 2…..N. The two basis vectors are related by the vielbiens $e^A_{\alpha}$ through $e_{\alpha} = e_{\alpha}^A v_A$. $\mathbf{n}_{\hat{J}}$ constitue a set of $N-4$ orthonormal vectors, orthogonal
to the subspace spanned by $\mathbf{e}_{\alpha}$, that is, $\mathbf{n}_{\hat{J}}\cdot\mathbf{e}_{\alpha}=0$  and $\ \mathbf{n}_{\hat{J}}\cdot\mathbf{n}_{\hat{I}}\mathbf{=}\delta_{\hat{J}\hat{I}},$ where $\hat{J},\hat{I}=5,6,...,N$. The vectors $\mathbf{n}_{\hat{J}}$, $\mathbf{e}_{\alpha}$ form a complete basis in tangent space and therefore $\mathbf{v}_{A}$ can be expanded as
\begin{equation}
\mathbf{v}_{A}=v_{A}^{\alpha}\mathbf{e}_{\alpha}+n_{A}^{\hat{J}}%
\mathbf{n}_{\hat{J}}\mathbf{.} \label{13}%
\end{equation}

The affine connection $\Gamma_{\alpha \beta}^{\nu}$ and the spin connection $w_{\beta A}^B$ defines the parallel transport for coordinate basis and vielbiens respectively
 \begin{equation}
\mathbf{\nabla}_{\mathbf{e}_{\beta}}\mathbf{e}_{\alpha}\equiv\mathbf{\nabla
}_{\beta}\mathbf{e}_{\alpha}=\Gamma_{\alpha\beta}^{\nu}\mathbf{e}_{\nu
},\ \ \mathbf{\nabla}_{\beta}\mathbf{v}_{A}=-\omega_{\beta A}^{\quad
\ B}\mathbf{v}_{B}, \label{5a}%
\end{equation}
The 2 connections are related by the metricity condition \\
\begin{equation}
\partial_{\beta}e_{A\alpha}=-\omega_{\beta A}^{\quad\ B}e_{B\alpha}%
+\Gamma_{\alpha\beta}^{\nu}e_{A\nu}. \label{4c}%
\end{equation}
$R_{\alpha \beta}^{AB} (w)$ is the Riemann curvature tensor of the spin connection of the group SO(1,N-1) defined by\\
\begin{equation}
R^{AB}_{\alpha \beta}(w)= \partial_{\alpha} w_{\beta} ^{AB}  -  \partial_{\beta} w_{\alpha} ^{AB} + w_{\alpha} ^{AC} w_{\beta C } ^B -w_{\beta} ^{AC} w_{\alpha C} ^B  \label{6}%
\end{equation}
and $R^{\rho}_{\gamma \alpha \beta}(\Gamma)$ is that of affine connection of the group SO(1,3) defined by\\
\begin{equation}
R_{\,\,\,\gamma\alpha\beta}^{\rho}\left(  \Gamma\right)  =\partial_{\alpha
}\Gamma_{\beta\gamma}^{\rho}-\partial_{\beta}\Gamma_{\alpha\gamma}^{\rho
}+\Gamma_{\alpha\kappa}^{\rho}\Gamma_{\beta\gamma}^{\kappa}-\Gamma
_{\beta\kappa}^{\rho}\Gamma_{\alpha\gamma}^{\kappa}, \label{30}%
\end{equation}
 The spin connection curvature $R_{\alpha\beta}^{\quad AB}\left(  \omega\right) $ can be expressed in terms of the affine connection
$R_{\,\,\,\gamma\alpha\beta}^{\sigma}\left(  \Gamma\right)  $ by \\
\begin{equation}
R_{\alpha\beta}^{\quad AB}\left(  \omega\right)  =R_{\alpha\beta}^{\quad
AC}\left(  \omega\right)  n_{C}^{\hat{I}}n_{\hat{I}}^{B}+R_{\,\,\,\gamma
\alpha\beta}^{\rho}\left(  \Gamma\right)  e_{\rho}^{A}e^{B\gamma}. \label{32}%
\end{equation}

Next, it is shown that the first term on the right hand side of this equation can be entirely expressed in terms of the spin connections defining the parallel transport of vectors $\mathbf{n}_{\hat{J}}$ in the subspace of tangent space orthogonal to those part spanned by the four coordinate basis vectors $\mathbf{e}_{\alpha}.$ These connections are denoted by
$A_{\beta\hat{J}}^{\quad\ \hat{I}}$. Finally, after making a choice of gauge where the mixed components of the vielbien vanish, eq 11 can be written as\\

\begin{equation}
R^{AB}_{\alpha \beta}(w)= F_{\alpha \beta}^{\hat{J} \hat{I}}n^A_{\hat{J}} n^B_{\hat{I}}+ R_{\gamma \alpha \beta} ^{\rho} (\Gamma) e_{\rho}^A e^{B \gamma}  \label{6}%
\end{equation}
 $F_{\alpha \beta}^{\hat{J} \hat{I}}$ is the curvature tensor or field strength of the remaining SO(N-4) group\\
\begin{equation}
F_{\alpha \beta}^{\hat{I} \hat{J}} (A) = \partial_{\alpha} A_{\beta} ^{\hat{I} \hat{J}}  -  \partial_{\beta} A_{\alpha} ^{\hat{I}\hat{J}} + A_{\alpha}^{\hat{I}\hat{L}}A_{\beta \hat{L}}^{\hat{J}} -  A_{\beta}^{\hat{I}\hat{L}}A_{\alpha \hat{L}}^{\hat{J}}  \label{6}%
\end{equation}
The first term to the right hand side of eq 12 correspond to SO(N-4) gauge theory and the second one to SO(1,3) gravity theory.\\
The realistic group allows us to unify all known interactions in one family, the SO(1,13), and leads to Einstein gravity with the SO(10) gauge group. However, not entirely equivalent to the SO(10) grand unified theory.\\

Based on the results of Chamseddine and Mukhanov, there are two ways to proceed : We can start by proving that the Pontryagin density in 4-D space time of the larger group SO(N) splits into one corresponding to gauge theory and another for gravity. Then we relate each Pontryagin density to its corresponding Chern-Simons Term in 3-D space time.
The second way is to start directly in 3-D space time and prove that the Chern-Simons term for the larger groups splits into two terms one for gauge theory and the other for gravity.\\
\subsection{ Pontryagin densities}
The starting points for the construction of the Chern-Simons terms are objects called Chern-Pontryagin densities. On a 2n dimensional manifold, these are of the form:\\
\begin{equation}
P^{2n} \propto \epsilon^{\mu_1 \mu_2 ...\mu_{2n}} Tr (F_{\mu_1 \mu_2}.... F_{\mu_{2n-1} \mu_{2n}}) \label{5}%
\end{equation}
where F (field strength) is the curvature 2-form $(dA + A\wedge A)$ of some G-connection (G is the gauge group). These are gauge-invariant, closed, and their integral over the manifold M (compact, no boundary) is an integer which is a topological invariant. These sorts of invariants are examples of characteristic classes.\\
The Pontryagin density for gauge theory is\\
\begin{equation}
P_4 = -\frac{1}{16\pi^2} Tr(^{\ast} F^{\mu \nu} F_{\mu \nu}) 
\end{equation}
Where\\
\begin{equation}
F_{\mu \nu}^a = \partial_{\mu} A_{\nu}^a - \partial_{\nu} A_{\mu}^a  + f^{abc} A_{\mu b}A_{\nu c}
\end{equation}
and that for gravity is the ( Hirzebruch-Pontryagin) given by\\
\begin{equation}
 \ast RR = \frac{1}{2} \epsilon^{\mu \nu \alpha \beta } R_{\mu \nu \rho \sigma} R_{\alpha \beta}^{\rho \sigma}
\end{equation}


Using eq 12, we can write the Pontryagin density for the large group SO(N) in 4D is\\
 \begin{equation}
 \frac{1}{2} \epsilon^{\mu \nu \alpha \beta} R^{AB}_{\mu \nu} R_{\alpha \beta AB}= \frac{1}{2} \epsilon^{\mu \nu \alpha \beta} (F_{\mu \nu}^{\hat{J} \hat{I}}n^A_{\hat{J}} n^B_{\hat{I}}+ R_{\gamma \mu \nu} ^{\rho} e_{\rho}^A e^{B \gamma})(F_{\alpha \beta}^{\hat{K} \hat{L}}n_{\hat{J}A} n_{\hat{L}B}+ R_{\sigma \alpha \beta} ^{\delta} e_{\delta A} e_B^{\sigma})  \label{7}%
\end{equation} 
Using the relation $n_{\hat{J}}^A e^{\alpha}_A=0$ $\cite{ChMu}$, we deduce that the mixed terms vanish and we are left with the Pontryagin densities corresponding to gauge theory and gravity.\\
\begin{equation}
\frac{1}{2} \epsilon^{\mu \nu \alpha \beta} R^{AB}_{\mu \nu} R_{\alpha \beta AB}= \frac{1}{2} \epsilon^{\mu \nu \alpha \beta} F_{\mu \nu \hat{K} \hat{L}} F_{\alpha \beta}^{\hat{K} \hat{L}} + \frac{1}{2} \epsilon^{\mu \nu \alpha \beta} R^{AB}_{\mu \nu} R_{\alpha \beta AB} \label{8}%
\end{equation}

 We can deduce that in 4-D space-time, the Pontryagin densities for both gauge theory and gravity can be unified. \\
We are interested in showing the unification in 3-D space time. The Chern-Pontryagin densities in 4-D are the exterior derivatives of the Chern-Simons entities in 3-D. 
\begin{equation}
\begin{aligned}
P_4 = Tr (F \wedge F)= Tr ((dA+ A \wedge A)(dA + A \wedge A))\\
= Tr(d(AdA +A^3)) = Tr(dw_3), 
\end{aligned}
\end{equation}

where $Tr(A^4)=0$.\\

We have $I_{CS} =\int {w_3} $ in 3-D. Then\\
\begin{equation}
 \int_{M_4} P_4 = \int_{M_4} dw_3 = \int_{\partial M_4} w_3  \label{9}%
\end{equation}
Based on this, we can easily translate the 4-D unification $\int_{M_4} P_4 = \int_{M_4} P_4^{gauge} + \int_{M_4} P_4^ {gravity}$ into a 3-D one of the corresponding Chern-Simons terms\\
\begin{equation}
 \int_{M_3} w_3 = \int_{M_3} w_3^{gauge} + \int_{M_3} w_3^{gravity}  \label{10}%
\end{equation}
Where\\
\begin{equation}
\begin{aligned}
  w_3^{gauge} = \epsilon^{ijk}(A_i^a \partial_j A_k^a + \frac{1}{3} f^{abc} A_i^a A_j^b A_k^c) \\
	 w_3^{gravity} = \epsilon^{ijk} (R_{ij ab} w_{k}^{ab} + \frac{2}{3} w_{i b}^c w_{j c}^a w_{k a}^b) \label{11}%
	\end{aligned}
\end{equation}
\subsection{ 3-D Chern-Simons terms}
Alternatively, we can show the unification by starting directly in 3-D space-time. The topological Chern-Simons term corresponding to the larger group SO(N) is\\
 \begin{equation}
I_{CS}= \frac{k}{4 \pi} \int  Tr ( A \wedge dA + \frac{2}{3} A \wedge A \wedge A), \label{12}%
\end{equation} 
Where A is the 1-form connection of the group, $A= dx^{\mu} \frac{1}{4} A_{\mu}^{AB} \Gamma_{AB}$ and A,B,C range from 1 to N.\\
$\Gamma$-matrices is a set of $2^D$ matrices resulting from the repeated multiplication of the $\gamma$-matrices of the D-dimensional clifford algebra given by $\cite{DeWitt}$.\\
\begin{equation}
\gamma_a \gamma_b + \gamma_b \gamma_a = 2 \delta_{ab} I 
\end{equation} 
Where a,b ranges from 1 to D.\\
Following the properties of $\Gamma$ matrices in an arbitrary space-time dimension  $\cite{DeWitt}$, the Chern-Simons term becomes\\
\begin{equation}
I_{CS}= \frac{k}{4 \pi} \int (A^{AC} dA^{CA} + \frac{2}{3} A^{AB} A^{BC} A^{CA}) \label{13}
\end{equation}     
Now, we can split the indices A, B, C into a, b, c which go from 1 to 3 and $\hat{I}, \hat{J}, \hat{K}$ which ranges from 4 to N. We get\\
\begin{equation}
\begin{aligned}
I_{CS}= \frac{k}{4 \pi} \int A^{AC} dA^{CA} + \frac{2}{3} A^{AB} A^{BC} A^{CA}= \\
 \frac{k}{4 \pi} \int A^{ab} dA^{ba} + \frac{2}{3} A^{ab} A^{bc} A^{ca} +
\frac{k}{4 \pi} \int A^{\hat{I} \hat{J}} dA^{\hat{J} \hat{I}} + \frac{2}{3} A^{\hat{I} \hat{J}} A^{\hat{J} \hat{K}} A^{\hat{K} \hat{I}} \label{14}%
\end{aligned}
\end{equation}  		
We notice that the first term in eq(27) corresponds to gravity Chern-Simons action and the second one to gauge Chern-Simons action.\\

\section{Consequences of this Unification}
\subsection{Quantization of the Coefficient}
The coefficient k of the Chern-Simons term for compact simple groups is quantized $\cite{Jones}.$ To unify non-abelian Chern-Simons gauge theory with the gravity Chern-Simons theory, we have two possibilities. For space-time with Euclidean signature, we can choose our group to be the compact SO(6) group which will split into SO(3) for gravity term and SO(3) for gauge theory term. Such choice leads to the quantization of the coulping constant k for the three terms, which must be the same. Thus, we get the quantization of the coupling constant of the Gravity Chern-Simons term with the compact group SO(3). The other possibility is for manifolds with Lorentzian signature. We choose the non-compact group SO(1,5) which splits into SO(1,2) for gravity and SO(3) for gauge theory. The homotopy group of SO(1,5) is equal to that of SO(5)\\
\begin{equation}
  \pi_5(SO(1,5)) = \pi_5(SO(5)) = Z_2, 
\end{equation}
the coefficient k must not be quantized as the winding number is not sensitive to torsion and vanishes $\cite{topological}$.\\ 

\subsection{Topological Invariants}
A topological Field Theory is a theory that computes toplogical invariants. Witten $\cite{Jones}$ proved that for the weak coupling limit (large k), the partition function\\
 \begin{equation}
Z = \int DA exp (\frac{ik}{4 \pi} \int_{M} Tr( AdA + \frac{2}{3}A \wedge A \wedge A) \label{15}%
\end{equation}
is a toplogical invariant quantity of the 3-D space-time manifold. We are going to proceed as Witten did. For the SO(6) group, the partition function Z splits into two parts\\
 \begin{equation}
\begin{aligned}
Z = \int DA exp( iI_{CS}) = \int DA exp(iI_{gauge} + iI_{gravity}) = \\
\int Dw exp(iI_{gravity}) \int DB exp( iI_{gauge})=Z_1 . Z_2  
\end{aligned}
\end{equation}
Where the first term is the SO(1,2) Gravity Chern-Simon term with w as a gauge connection and the second term is the SO(3) gauge Chern-Simon term with B as a gauge field.\\
We consider the weak coupling limit of the gauge part. The weak coupling limit of $Z_2$ is given by\\
 \begin{equation}
Z_2 = \sum_{\alpha} \mu(B^{\alpha})
 \end{equation} 
Where $\mu(B^{\alpha})$ is a function of flat connections for which the curvature vanishes. We expand the gauge field $B_i = B_i^{(\alpha)} + C_i$, the Chern-Simon gauge action term becomes\\
\begin{equation}
I_{CS}^{gauge} = k I(B^{\alpha})+ \frac{k}{4\pi}\int_M Tr(C \wedge DC) 
 \end{equation}  
Where D is the covariant derivative with respect to $B^{\alpha}$.\\
To carry out the gaussian integral in eq(32), a gauge fixing is needed which can not be done without picking a metric. We choose such a metric to satisfy $D_i C^i =0$. The resulting ghost action becomes\\
\begin{equation}
S_{GF} = \int_M Tr(\phi D_i C^i + \bar{c} D_i D^i c)  
 \end{equation}
Where $\phi$ is a lagrangian multiplier that enforces the gauge condition $D_i C^i=0 $ and c,$\bar{c}$ are anticommuting ghosts.\\
Integrating out C, $\phi$, c, $\bar{c}$, we get\\
\begin{equation}
exp(\frac{i \pi \eta(B^{\alpha})}{2}) T_{\alpha}
\end{equation}
Where $\eta(B^{\alpha})$ is the " eta inavraint" defined by\\

\begin{equation}
\eta(B^{\alpha}) = \frac{1}{2} \lim_{s\to 0} \sum_i sign \lambda_i |\lambda_i|^{-s}
\end{equation}
Where $\lambda_i$'s are eigenvalues of operator $L_i$, the restriction of $\ast D_B + D_B \ast$ on odd forms, $T_{\alpha}$ is the torsion invariant of flat connection $B^{(\alpha})$. Using Atiyah-Patodi-Singer index theorem, the partition function can be written as\\
\begin{equation}
Z_2 =  exp (i \frac{\pi}{2} \eta (0)) \sum_{\alpha} e^{i(k+c_2(G)/2)I(B^{(\alpha)})}. T^{\alpha} \label{15}%
\end{equation}
$\eta (0)$ is the eta invariant of the trivial gauge field and $c_2(G)$ is the Casimir operator of G. $\eta (0)$  is the only term that depends on the metric. Witten suggested that by adding a counter term, the partition function will be regualrized and turned  into a topological invariant quantity. This counter term must be a multiple of the gravitational Chern-Simons term. This term is already present in our case. Substitute the weak coupling limit of $Z_2$ eq(36) in eq (30) we get\\
 \begin{equation}
\begin{aligned}
Z =\int Dw exp(i I_{grav}).Z_2\\
=\int Dw exp(i I_{grav}) exp (i \frac{\pi}{2} \eta (0)) \sum_{\alpha} e^{i(k+c_2(G)/2)I(B^{(\alpha)})}. T^{\alpha}\\
=\int Dw exp(i (I_{grav} + \frac{\pi}{2} \eta (0)))  \sum_{\alpha} e^{i(k+c_2(G)/2)I(B^{(\alpha)})} T^{\alpha}
\end{aligned}
\end{equation}

The added gravitational Chern-Simons term by Witten is already present in our case. So our partition function is a toplogical invariant quantity without the adding any term.\\
\bigskip

Summarizing, we have shown how gravity and gauge Chern-Simons theories could be unified in 3-D space-time. The coupling constant is quantized for compact groups but not for non-compact ones.  Moreover, the partition function Z directly becomes a topological invariant quantity without the need of adding any term as Witten did  $\cite{Jones}$. It should be noted that such Chern-Simon theories admit local supersymmetric extensions. This is achieved by considering the gauging of supergroups, or by extending the space-time manifold to a supermanifold, or both $\cite{topological}$. We shall consider the graded groups that are extensions of our SO(1,5) group. Referring to $\cite{topological}$, we could choose the supergroup\\
\begin{equation}
O(6,1) \oplus SU(2),(8,2)
\end{equation}
Where we can prove that the unification still exist for supersymmetric Chern-Simon theories. Details of this will be dealt with in the future.

\

\bigskip

\textbf{{\large {Acknowledgments}}}

We would like to thank Professor Ali Chamseddine for suggesting the problem and for his helpful discussions on the subject. We would like to thank also the American University of Beirut (Faculty of Science) for the support.
.\bigskip\ 

\bigskip

\bigskip
\end{document}